\documentclass[pre,showpacs,showkeys,preprintnumbers,amsmath,amssymb,superscriptaddress,twocolumn]{revtex4}
\usepackage{graphicx}
\usepackage{amsfonts}
\usepackage{url}
\usepackage{epstopdf}
\usepackage{color}
\usepackage{tikz-cd}
\usepackage{bm}
\usepackage[colorlinks=true,linkcolor=blue,citecolor=red]{hyperref}%

\def\var{{\rm var}}

\def\mean{{\rm mean}}

\def\R{{\mathbb R}}

\def\M{{\bf M}}
\def\Pq{{\bf P}}
\def\V{{\bf V}}
\def\D{{\bf D}}
\def\K{{\bf K}}

\def\x{\bm{x}}

\begin{document}

\title{Time-averaged MSD for switching diffusion}

\author{Denis~S.~Grebenkov}
 \email{denis.grebenkov@polytechnique.edu}
\affiliation{
Laboratoire de Physique de la Mati\`{e}re Condens\'{e}e (UMR 7643), \\ 
CNRS -- Ecole Polytechnique, University Paris-Saclay, 91128 Palaiseau, France}

\date{\today}

\begin{abstract}
We consider a classic two-state switching diffusion model from a
single-particle tracking perspective.  The mean and the variance of
the time-averaged mean square displacement (TAMSD) are computed
exactly.  When the measurement time (i.e., the trajectory duration) is
comparable to or smaller than the mean residence times in each state,
the ergodicity breaking parameter is shown to take arbitrarily large
values, suggesting an apparent weak ergodicity breaking for this
ergodic model.  In this regime, individual random trajectories are not
representative while the related TAMSD curves exhibit a broad spread,
in agreement with experimental observations in living cells and
complex fluids.  Switching diffusions can thus present, in some cases,
an ergodic alternative to commonly used and inherently non-ergodic
continuous-time random walks that capture similar features.
\end{abstract}

%02.50.-r       (Probability theory, stochastic processes, and statistics)
%05.40.-a 	Fluctuation phenomena, random processes, noise, and Brownian motion
%02.70.Rr       (General statistical methods)
%05.10.Gg 	Stochastic analysis methods (Fokker-Planck, Langevin, etc.) 

%02.50.Ey 	Stochastic processes  (Probability theory, stochastic processes, and statistics)


\maketitle

\section{Introduction}

A continuous-time random walk (CTRW) was originally proposed to model
hopping processes in semiconductors with randomly distributed heavy
tailed waiting times due to random energetic trapping
\cite{Montroll65,Bouchaud90}.  This model has been extensively studied
and rapidly became an archetypical model of anomalous diffusion in
different contexts, including the intracellular transport in
microbiology \cite{Metzler00,Bressloff13,Metzler14}.  This is also an
emblematic model of weak ergodicity breaking (WEB) and aging phenomena
\cite{Bouchaud92,Burov11,Schulz13}.  In fact, when the waiting time
distribution has no first moment, there is no characteristic time
scale for waiting events, so that the duration of a single stalling
period may be comparable to the overall measurement time that
prohibits self-averaging needed for ergodic behavior.  Moreover,
longer measurement times favor longer waiting times so that the
statistical properties of the system depend on the measurement time.
In the biological context, CTRWs were applied to mimic molecular
caging in an overcrowded intracellular environment when a diffusing
macromolecule, surrounded by other macromolecules and filaments, may
wait a relatively long time in such an effective molecular ``cage''
before being able to jump to a next cage
\cite{Wong04,Weigel11,Jeon11,Barkai12}.  However, the validity of the
considered heavy-tailed distribution of waiting times in the
biological context remains debatable \cite{Szymanski09}.  In
particular, such a distribution may have an exponential cut-off so
that anomalous features are transient, before getting into
normal diffusion at long times \cite{Burov11,Lanoiselee16}.  Although
a CTRW with cut-off sounds more realistic, the underlying mathematical
formalism is rather difficult.

In this light, a simpler model of random switching between several
diffusion states, characterized by diffusion coefficient
$D_1,\ldots,D_J$, is appealing.  A particle started in a state $i$
undergoes normal diffusion with diffusivity $D_{i}$ for a random time,
until it switches to another state $j$, with the switching rate
$k_{ij}$, and so on.  Diffusion states can represent different
conformations of a large macromolecule (e.g., globular versus
filamentous structures) and thus distinct effective radii.
Alternatively, diffusion states can account for eventual temporal
binding of the diffusing particle to other molecules that may either
slow down or even stop its motion.  Such switching diffusions are
often employed to describe the dynamics in biological systems
\cite{Bressloff17,Sungkaworn17,Weron17} and used as simple models of
intermittent processes \cite{Lanoiselee17}.

In this paper, we mainly focus on the two-state model with diffusion
coefficients $D_1$ and $D_2$ and the switching rates $k_{12}$ and
$k_{21}$.  Note that $1/k_{12}$ and $1/k_{21}$ are the mean residence
times of the particle in the states $1$ and $2$, respectively. The
molecular caging effect can be modeled by setting $D_2 = 0$, i.e., the
particle does not move in the second state.  This model has been
extensively studied, in particular, in the nuclear magnetic resonance
literature, in which it is known as the K\"arger model
\cite{Karger85,Fieremans10,Moutal18}.  Here, we look at this model from a
single-particle tracking perspective.  This simple model will allow us
to investigate the reproducibility of measurements over individual
random trajectories and the effect of their duration $T$.  In
particular, if both residence times $1/k_{12}$ and $1/k_{21}$ are
small as compared to $T$, the particle switches very often between two
states and manages to probe these states reliably during the
measurement time.  In this limit, the intermittent process is seen as
an ergodic normal diffusion with some mean diffusivity $\bar{D}$ (see
below).  In contrast, if both residence times are much larger than
$T$, the particle remains within a single state over the whole
measurement with a high probability.  As a consequence, such a single
trajectory would bring information only about one state, whereas
another trajectory may bring information only about the other state.
In other words, individual trajectories are not representative of the
whole dynamics and thus, from a practical point of view, not ergodic.
Since the switching model is ergodic, we call this regime {\it
apparent} WEB due to insufficient measurement time.  Finally, the
situation when one or both residence times are comparable to the
measurement time is a borderline case.  This model will thus help us
to study the transition from one regime to the other that is quite
typical for intracellular transport measurements.

In order to quantify the reliability of measurements and an eventual
apparent WEB, we investigate the time-averaged mean square
displacement (TAMSD) of a particle undergoing such an intermittent
motion:
\begin{equation}  \label{eq:TAMSD}
\overline{\delta^2}(t,T) = \frac{1}{T-t} \int\limits_0^{T-t} dt_0 \, (X(t+t_0) - X(t_0))^2 ,
\end{equation}
where $t$ is the lag time.  We compute exactly both the mean and the
variance of the random variable $\overline{\delta^2}(t,T)$ to evaluate
the ergodicity breaking (EB) parameter $\chi$, also known as the
squared coefficient of variation of the TAMSD \cite{He08,Burov10}:
\begin{equation}  \label{eq:chi_def}
\chi = \frac{\langle [\overline{\delta^2}(t,T)]^2 \rangle}{\langle \overline{\delta^2}(t,T) \rangle^2} - 1 
= \frac{\var\{\overline{\delta^2}(t,T)\}}{(\mean\{\overline{\delta^2}(t,T)\})^2} \,,
\end{equation}
where $\langle \ldots \rangle$ denotes the expectation, i.e., the
ensemble average over the space of all possible trajectories $X(t)$ of
duration $T$.

The paper is organized as follows.  In Sec. \ref{sec:theory}, we
provide the propagator for multiple successive points in a matrix form
which is then used to derive the mean and the variance of the TAMSD.
Different asymptotic regimes of the resulting EB parameter are then
analyzed.  In Sec. \ref{sec:discussion}, we discuss an apparent WEB
when the measurement time is comparable to or smaller than the mean
residence times.  In particular, we illustrate a large spread of TAMSD
curves obtained via numerical simulations.  We also mention extensions
for multi-state switching diffusion and restricted diffusion in
bounded domains.  Section \ref{sec:conclusion} summarizes the results
and concludes.

\section{Theoretical results}
\label{sec:theory}

\subsection{Propagator}

We consider a particle that diffuses on a real axis $\R$ and switches
randomly between two diffusivities $D_1$ and $D_2$ with rates $k_{12}$
and $k_{21}$ (see Sec. \ref{sec:extension1} and \ref{sec:extension2}
for extensions).  Let $P_{ii_0}(x,t|x_0,t_0)$ be the propagator of
this particle, i.e., the probability density of finding the particle
at point $x$ in state $i$ at time $t$ given that it started from point
$x_0$ in state $i_0$ at time $t_0$.  These four propagators satisfy
four coupled partial differential equations
\begin{subequations}  \label{eq:coupled}
\begin{eqnarray}
\frac{\partial}{\partial t} P_{1,i_0} &=& D_1 \frac{\partial^2}{\partial x^2} P_{1,i_0} - k_{12} P_{1,i_0} + k_{21} P_{2,i_0} ,   \\
\frac{\partial}{\partial t} P_{2,i_0} &=& D_2 \frac{\partial^2}{\partial x^2} P_{2,i_0} - k_{21} P_{2,i_0} + k_{12} P_{1,i_0} 
\end{eqnarray}
\end{subequations}
(with $i_0 = 1,2$), subject to the initial conditions:
$P_{i,i_0}(x,t=t_0|x_0,t_0) = \delta_{i,i_0} \delta(x-x_0)$ (a
rigorous mathematical formulation of switching models and of some
their properties can be found in \cite{Yin10,Yin10b,Baran13}).

The solution of these equations is well known (e.g., see
\cite{Karger85}).  Here we recall the main formulas that will be
needed for the analysis of the TAMSD.  In the Fourier space,
\begin{equation}
P_{ii_0}(x,t|x_0,t_0) = \int\limits_{-\infty}^{\infty} \frac{dq}{2\pi} \, e^{-iq(x-x_0)} \, \hat{P}_{ii_0}(q,t-t_0),
\end{equation}
Eqs. (\ref{eq:coupled}) are reduced to coupled first-order
differential equations whose solution is easily found in a matrix form
\begin{equation}
\Pq(q,t) = \left(\begin{array}{c c} \hat{P}_{11}(q,t) & \hat{P}_{12}(q,t) \\ \hat{P}_{21}(q,t) & \hat{P}_{22}(q,t) \\ \end{array} \right)
=  \exp(- \M_q t),
\end{equation}
where
\begin{equation}
\M_q = \left(\begin{array}{c c} D_1 q^2 + k_{12} & - k_{21} \\ - k_{12} & D_2 q^2 + k_{21} \\  \end{array} \right).
\end{equation}
The eigenvalues and eigenvectors of the matrix $\M_q$, 
\begin{equation*}
\M_q \V_q = \V_q \left(\begin{array}{c c} \gamma_q^{+} & 0 \\ 0 & \gamma_q^{-} \\  \end{array} \right).
\end{equation*}
are known explicitly:
\begin{eqnarray}
\gamma_q^\pm &=& \frac12 \biggl((D_1+D_2)q^2 + (k_{12} + k_{21}) \\  \nonumber
&\pm& \sqrt{((D_2-D_1)q^2 + (k_{21}-k_{12}))^2 + 4k_{12} k_{21}} \biggr)
\end{eqnarray}
and
\begin{equation}
\V_q = \left(\begin{array}{c c} \displaystyle \frac{k_{21}}{k_{12} + D_1 q^2 - \gamma^{+}_q} & 
 \displaystyle  \frac{k_{21}}{k_{12} + D_1 q^2 - \gamma^{-}_q} \\ 1 & 1 \\  \end{array} \right).
\end{equation}
One gets thus
\begin{equation}
\Pq(q,t) = \frac{
\left(\begin{array}{c c} 
e^{-\gamma_q^{+} t} \mu_q^{-} - e^{-\gamma_q^{-} t} \mu_q^{+} & k_{21}(e^{-\gamma_q^{-} t} - e^{-\gamma_q^{+} t}) \\ 
k_{12}(e^{-\gamma_q^{-} t} - e^{-\gamma_q^{+} t}) & e^{-\gamma_q^{-} t} \mu_q^{-} - e^{-\gamma_q^{+} t} \mu_q^{+} \\ \end{array} \right)
}{\gamma_q^{+} - \gamma_q^{-}} ,
\end{equation}
with $\mu_q^{\pm} = D_1q^2 + k_{12} - \gamma_q^{\pm}$.  In the special
case $q = 0$, one has $\gamma_0^{+} = k_{12} + k_{21}$ and
$\gamma_0^{-} = 0$.

To compute the moments of the TAMSD, one needs the propagator for
multiple successive points, which is simply the product of the above
propagators due to the Markov property of the process:
\begin{eqnarray}  \nonumber
&& P(x_n,i_n,t_n;\, x_{n-1},i_{n-1},t_{n-1}; \ldots; x_1,t_1,i_1;\, x_0,i_0,0) \\  \nonumber
&& = P_{i_n,i_{n-1}}(x_n,t_n|x_{n-1},t_{n-1}) \ldots  P_{i_1,i_0}(x_1,t_1|x_0,0)  \\  \nonumber
&& = \int\limits_{\R^n} \frac{dq_1}{2\pi} \ldots \frac{dq_n}{2\pi} e^{-iq_n(x_n-x_{n-1}) - \ldots - iq_1(x_1 - x_0)} \\  \label{eq:multiple_prop1}   
&& \times \hat{P}_{i_n,i_{n-1}}(q_n,t_n-t_{n-1}) \ldots \hat{P}_{i_1,i_0}(q_1,t_1).
\end{eqnarray}
Denoting by $p_1$ (resp., $p_2 = 1 - p_1$) the probability of starting
in the state $1$ (resp. 2) at time $0$, the marginal propagator
averaged over the state variables $i_k$ reads
\begin{eqnarray}  \nonumber
&& P(x_n,t_n;\, x_{n-1},t_{n-1}; \ldots; x_1,t_1;\, x_0,0) \\  \nonumber
&& = \int\limits_{\R^n} \frac{dq_1}{2\pi} \ldots \frac{dq_n}{2\pi} e^{-iq_n(x_n-x_{n-1}) - \ldots - iq_1(x_1 - x_0)} \\  \label{eq:multiple_prop2}   
&& \times {\mathcal P}_n(q_n,t_n-t_{n-1}; \ldots; q_2,t_2-t_1;\, q_1,t_1),
\end{eqnarray}
where
\begin{eqnarray}  \label{eq:Pmultiple}
&& {\mathcal P}_n(q_n,t_n-t_{n-1}; \ldots; q_2,t_2-t_1;\, q_1,t_1) \\  \nonumber
&& = \left(\begin{array}{c} 1 \\ 1 \\ \end{array} \right)^\dagger
\Pq(q_n,t_n-t_{n-1})   \ldots  \Pq(q_1,t_1) \left(\begin{array}{c} p_1 \\ p_2 \\ \end{array} \right).
\end{eqnarray}

\subsection{Mean TAMSD}

We first calculate the characteristic function of the displacement
between times $t_1$ and $t_2$ such that $0 < t_1 < t_2$:
\begin{eqnarray}  \nonumber
G(q) &\equiv& \langle e^{iq(X(t_2)-X(t_1))} \rangle \\  \nonumber
&=& \int\limits_{-\infty}^\infty dx_1 \int\limits_{-\infty}^\infty dx_2 \,
e^{iq(x_2 - x_1)} \, P(x_2,t_2; x_1,t_1; x_0,0) \\
&=& {\mathcal P}_2(q,t_2-t_1;\, 0,t_1) .
\end{eqnarray}
From this characteristic function, one can compute the moments of the
displacement, in particular, the mean square displacement:
\begin{eqnarray}  \label{eq:auxil1}
&& \langle (X(t+t_0) - X(t_0))^2\rangle = - \lim\limits_{q\to 0} \frac{\partial^2 G(q)}{\partial q^2}  \\  \nonumber
&& = 2 \bar{D} t + 2 \frac{(p_1 k_{12} - p_2 k_{21})(D_1 - D_2)}{k^2} (1 - e^{-kt}) e^{-kt_0} \,, 
\end{eqnarray}
where
\begin{equation}
k = k_{12} + k_{21} 
\end{equation}
and
\begin{equation}
\bar{D} = \frac{D_1 k_{21} + D_2 k_{12}}{k} 
\end{equation}
is the mean diffusivity.
In particular, the ensemble averaged MSD is
\begin{equation}
\langle X^2(t)\rangle = 2 \bar{D} t + 2 \frac{(p_1 k_{12} - p_2 k_{21})(D_1 - D_2)}{k^2} (1 - e^{-kt}) \,. 
\end{equation}
From Eq. (\ref{eq:auxil1}), one can also deduce the mean TAMSD
\begin{eqnarray}
&& \langle \overline{\delta^2}(t,T) \rangle = 2 \bar{D} t  \\  \nonumber
&&+ \frac{2(p_1 k_{12} - p_2 k_{21})(D_1 - D_2)}{k^3(T-t)} (1 - e^{-kt}) (1 - e^{-k(T-t)}) \,. 
\end{eqnarray}

If the initial probabilities $p_1$ and $p_2$ are set to be from the
equilibrium, 
\begin{equation}  \label{eq:p_ini}
p_1 = p_1^{\rm eq} = \frac{k_{21}}{k},  \qquad  
p_2 = p_2^{\rm eq} = \frac{k_{12}}{k} ,
\end{equation}
then
\begin{equation}  \label{eq:MSD_equil}
\langle \overline{\delta^2}(t,T) \rangle = 2\bar{D} t .
\end{equation}
One cannot therefore reveal the intermittent character of this process
from the mean TAMSD alone.  Moreover, the mean value does not depend
on the measurement time $T$, as for normal Brownian motion.

\subsection{Variance of TAMSD}

The computation of the variance of the TAMSD is much more involved as
it requires the four-point correlation function.  In fact, one has
\begin{eqnarray}  \label{eq:auxil3}
\langle [\overline{\delta^2}]^2 \rangle &=& \frac{2}{(T-t)^2} \int\limits_0^{T-t} dt_0 \int\limits_{t_0}^{T-t} dt'_0 \\  \nonumber
&\times&\langle (X(t_0+t)-X(t_0))^2 (X(t'_0+t)-X(t'_0))^2 \rangle ,
\end{eqnarray}
from which the variance follows as usual:
\begin{equation}
\var\{ \overline{\delta^2} \} = \langle [\overline{\delta^2}]^2 \rangle - \langle \overline{\delta^2} \rangle^2 .
\end{equation}

For computing the second moment of the TAMSD, we introduce the
characteristic function
\begin{equation}
G(q,q') \equiv \langle e^{iq(X(t_0+t)-X(t_0)) + iq'(X(t'_0+t)-X(t'_0))} \rangle  .
\end{equation}
Since we have $t_0 < t'_0$ in Eq. (\ref{eq:auxil3}), there are two
cases:

(i) for $t_0 < t_0 + t < t'_0 < t'_0 + t$, we get
\begin{equation}
G(q,q') = {\mathcal P}_4(q', t;\, 0,t_0'-t_0-t;\, q,t;\, 0,t_0) ;
\end{equation}

(ii) for $t_0 < t'_0 < t_0 + t < t'_0 + t$, we get
\begin{equation}  \label{eq:auxil4}
G(q,q') = {\mathcal P}_4(q', t_0'-t_0;\, q+q',t_0+t-t_0';\, q,t_0'-t_0;\,  0,t_0) .
\end{equation}

In the evaluation of the integral in Eq. (\ref{eq:auxil3}), we
consider separately two cases: $t < T/2$ and $t > T/2$.

\subsubsection*{Case $t < T/2$}

In this case, one can split the integral in Eq. (\ref{eq:auxil3}) into
three parts
\begin{align}  \label{eq:auxil32}
& \langle [\overline{\delta^2}]^2 \rangle = \frac{2}{(T-t)^2} \biggl\{\int\limits_0^{T-2t} dt_0 \int\limits_{0}^{t} dt'_0 \, F_2(t,t_0,t'_0)  \\  \nonumber
& + \int\limits_0^{T-2t} dt_0 \int\limits_{0}^{T-2t-t_0} dt'_0 \, F_1(t,t_0,t'_0)  \\  \nonumber
& + \int\limits_0^t dt_0 \int\limits_{0}^{t_0} dt'_0 \, F_2(t,T-t-t_0,t'_0)  \biggr\} 
\end{align}
(note that the integration variables $t_0$ and $t'_0$ were shifted),
with
\begin{eqnarray*}
F_1(t,t_0,t'_0) &=& \lim\limits_{\substack{q\to 0\\ q'\to 0}} \frac{\partial^4 {\mathcal P}_4(q', t ;\, 0,t_0'; \, q,t ;\, 0,t_0)}
{\partial q^2 \, \partial q'^2} \, , \\
F_2(t,t_0,t'_0) &=& \lim\limits_{\substack{q\to 0\\ q'\to 0}} \frac{\partial^4 {\mathcal P}_4(q', t_0'; q+q',t-t_0';  q,t_0';  0,t_0)}
{\partial q^2 \, \partial q'^2} \,.
\end{eqnarray*}
After long and cumbersome computations of these derivatives and
integrals in Eq. (\ref{eq:auxil32}), we derive the following
expression for the variance of the TAMSD for $t < T/2$:
\begin{widetext}
\begin{eqnarray}  \label{eq:varchi1}
\var\{\overline{\delta^2}\} &=& \frac{4\bar{D}^2 t^3(4T - 5t)}{3(T-t)^2} + \frac{8k_{12} k_{21} (D_1-D_2)^2}{k^6(T-t)^2}  
\biggl\{ e^{-kT}(1 - e^{kt})^2 - 2(3 + 2k(T-t)) e^{-kt} \\  \nonumber
&+& k T (3(kt)^2 - 4kt + 4) - 2(2(kt)^3 - 3(kt)^2 + 5 kt - 3)\biggr\} ,
\end{eqnarray}
\end{widetext}
% see computation in 'karger_other_variance.mw'
where the initial probabilities $p_i$ were set to their equilibrium
values $p_i^{\rm eq}$ in Eq. (\ref{eq:p_ini}) to get a more compact
expression.

\subsubsection*{Case $t > T/2$}

In this case, only the second option in Eq. (\ref{eq:auxil4}) is
possible so that one does not need to split the integral in
Eq. (\ref{eq:auxil3}) into three parts, and one gets
\begin{equation}
\langle [\overline{\delta^2}]^2 \rangle = \frac{2}{(T-t)^2} \int\limits_0^{T-t} dt_0 \int\limits_{0}^{T-t-t_0} dt'_0 \, F_2(t,t_0,t'_0) ,
\end{equation}
where the integration variable was shifted in the second integral.
The evaluation of this integral yields the variance of the TAMSD for
$t > T/2$:
\begin{widetext}
\begin{eqnarray}  \label{eq:varchi2}
\var\{\overline{\delta^2}\} &=& \frac{4\bar{D}^2 (T^2 - 6Tt + 11t^2)}{3} + \frac{8k_{12} k_{21} (D_1-D_2)^2}{k^6(T-t)^2}  
\biggl\{ e^{-kT}(1 - 2e^{kt}) - 2(3 + 2k(T-t)) e^{-kt}  \\  \nonumber
&+& 5 e^{k(T-2t)} - \biggl(k^3T^3 - 2k^2T^2(3kt-1) + kT(9k^2t^2 - 4kt + 2) - 2(2k^3t^3 - k^2t^2 + kt + 1)\biggr) \biggr\} ,
\end{eqnarray}
\end{widetext}
where the initial probabilities $p_i$ were again set to their
equilibrium values $p_i^{\rm eq}$ for simplicity.

\subsection{Ergodicity breaking parameter}

The expressions (\ref{eq:varchi1}, \ref{eq:varchi2}) for the variance
of the TAMSD present the main computational result of this paper.  The
first term in these expressions is the variance of the TAMSD for
Brownian motion with the mean diffusivity $\bar{D}$
\cite{Qian91,Grebenkov11,Andreanov12,Grebenkov13}.  One can check that
the second term in Eqs. (\ref{eq:varchi1}, \ref{eq:varchi2}) is
nonnegative, i.e., the switching between two states can only increase
the variance of the TAMSD.

As Eqs. (\ref{eq:varchi1}, \ref{eq:varchi2}) for the variance are
provided exclusively for the case $p_i = p_i^{\rm eq}$, for which the
mean TAMSD in Eq. (\ref{eq:MSD_equil}) is particularly simple, the
analysis of the variance is equivalent, up to a simple multiplicative
factor $(2\bar{D} t)^2$, to that of the ergodicity breaking parameter
$\chi$ defined by Eq. (\ref{eq:chi_def}).  In the rest of the paper,
we focus on this parameter.

As the trajectory duration $T$ goes to infinity (for a fixed lag time
$t$), the EB parameter vanishes asymptotically as
\begin{eqnarray}  \label{eq:chi_leading}
\chi &\simeq & T^{-1} \biggl(\frac{4t}{3} + \frac{2k_{12} k_{21}(D_1-D_2)^2}{k^3 \bar{D}^2} \\  \nonumber
& \times & \frac{3(kt)^2-4kt+4-4e^{-kt}}{(kt)^2} \biggr) + O(T^{-2}),
%\var\{\overline{\delta^2} \} &\simeq & T^{-1} \biggl(\frac{16 \bar{D}^2 t^3}{3} + \frac{8k_{12} k_{21}(D_1-D_2)^2}{k^5} \\  \nonumber
%& \times & \bigl(3(kt)^2-4kt+4-4e^{-kt}\bigr) \biggr) + O(T^{-2}),
\end{eqnarray}
so that the switching process is ergodic, as expected.  In contrast,
in the double limit $k_{21} = p_1^{\rm eq} k \to 0$ and $k_{12}
= p_2^{\rm eq} k \to 0$ (with fixed $p_i^{\rm eq}$ and $k\to 0$),
Eq. (\ref{eq:varchi1}) yields
\begin{equation}  \label{eq:chi_ksmall} 
\chi \simeq \frac{p_1^{\rm eq} \, p_2^{\rm eq} (D_1-D_2)^2}{(p_1^{\rm eq} D_1 + p_2^{\rm eq} D_2)^2} + O(T^{-1}) .
\end{equation}
In this limit, the particle stays infinitely long in either of two
states, i.e., the process is not ergodic, and the variance of the
TAMSD does not vanish in the limit $T\to\infty$.  This singular
situation can also describe two populations of particles with distinct
diffusivities $D_1$ and $D_2$, and the leading term of the EB
parameter in Eq. (\ref{eq:chi_ksmall}) is a consequence of their
mixture with relative fractions $p_1^{\rm eq}$ and $p_2^{\rm eq}$.
This limit could alternatively be obtained by setting
$\overline{\delta^2} = \alpha \, \zeta_1 + (1-\alpha) \, \zeta_2$,
where $\zeta_i$ is the TAMSD for the $i$-th population (that differs
by the factor $D_i$), and a random selection between two populations
is realized by a Bernoulli random variable $\alpha$ taking the value
$1$ with probability $p_1^{\rm eq}$ and $0$ with probability $p_2^{\rm
eq} = 1 - p_1^{\rm eq}$.
Moreover, the EB parameter $\chi$ remains close to the limiting
expression (\ref{eq:chi_ksmall}) when $T \ll 1/k$.  In other words, if
the measurement time $T$ is short as compared to the residence time
$1/k$, the system exhibits an {\it apparent} WEB.  Clearly, the order
of two limits, $k_{12},\, k_{21} \to 0$ and $T\to\infty$, does matter
here: sending $T\to\infty$ for fixed $k_{12}$ and $k_{21}$ yields the
zero variance, as expected.

In the limit $t\to 0$, Eq. (\ref{eq:varchi1}) gives
\begin{equation}
\chi  \simeq \frac{2k_{12} k_{21} (D_1-D_2)^2 (kT - 1 + e^{-kT})}{k^4 \bar{D}^2 T^2} + O(t).
\end{equation}

\begin{figure}
\begin{center}
\includegraphics[width=85mm]{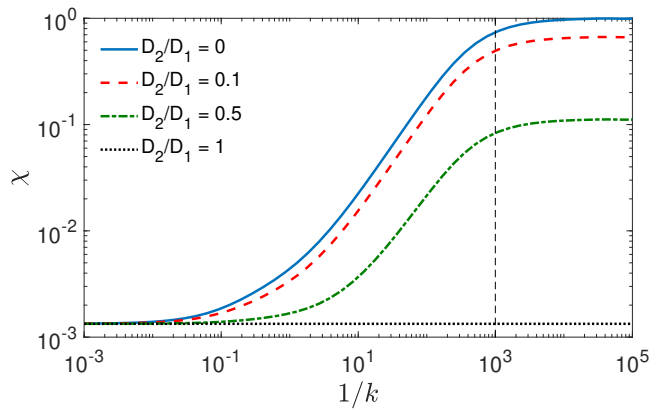} % chi_k2_.eps}
\end{center}
\caption{
(Color online) The EB parameter $\chi$ as a function of the residence
time $1/k$ for the case $k_{12} = k_{21} = k/2$, with $t = 1$, $T =
1000$, and several values of $D_2/D_1$ as indicated in the plot.
Arbitrary units are used.  Vertical dashed line indicates the
measurement time $T$.  Dotted horizontal line corresponding to normal
diffusion ($D_1 = D_2$) is close to $4t/(3T)$. }
\label{fig:kk}
% [C1,C2] = A_Karger_chivar_kk3;
\end{figure}

We first consider the particular case, in which two switching rates
are identical: $k_{12} = k_{21} = k/2$.  For our illustrative
purposes, we use dimensionless units for all parameters.  We set the
measurement time $T = 1000$ (i.e., the trajectory with a thousand
steps).  We recall that the EB parameter for Brownian motion is a
monotonously growing function of the lag time $t$, so that the
smallest available lag time $t = 1$ provides the most accurate
estimation of the TAMSD (see, e.g., \cite{Grebenkov11}).  One can
check that the same property holds for two-state switching diffusion.
As a consequence, we select the lag time $t = 1$ to be in the optimal
situation.  Figure \ref{fig:kk} shows the behavior of the EB parameter
as a function of the residence time $1/k$.  In the case of equal
diffusion coefficients, $D_2/D_1 = 1$, the states are identical, and
the switching model is reduced to normal diffusion.  The EB parameter
does not depend on the switching rate and is equal (in the leading
order) to $4t/(3T)$ according to Eq. (\ref{eq:chi_leading}).  In turn,
the stronger the difference between two states (i.e., smaller
$D_2/D_1$), the larger the EB parameter at large residence times
$1/k$.  In the extreme case of one immobile state (with $D_2 = 0$),
the EB parameter reaches the value $1$ according to
Eq. (\ref{eq:chi_ksmall}).  In contrast, a fast switching even between
very distinct states leads again to normal diffusion with the mean
diffusivity $\bar{D}$ and thus the EB parameter remains close to
$4t/(3T)$.  Note that in the considered case of equal switching rates,
the EB parameter does not exceed $1$.  We conclude that, due to slow
switching between states, the distribution of TAMSD even for the
smallest lag time ($t = 1$) can be relatively broad, i.e., the
standard deviation can be comparable to the mean value: $\chi \sim 1$.
In this situation, an increase of the trajectory duration $T$ can
improve the estimation only when $T$ exceeds the mean residence time
$1/k$.  This is in contrast with the case of normal diffusion, for
which the EB parameter is of the order of $t/T$ and can thus be made
very small, allowing for accurate estimations of the diffusion
coefficient for long enough trajectories.

\begin{figure}
\begin{center}
\includegraphics[width=85mm]{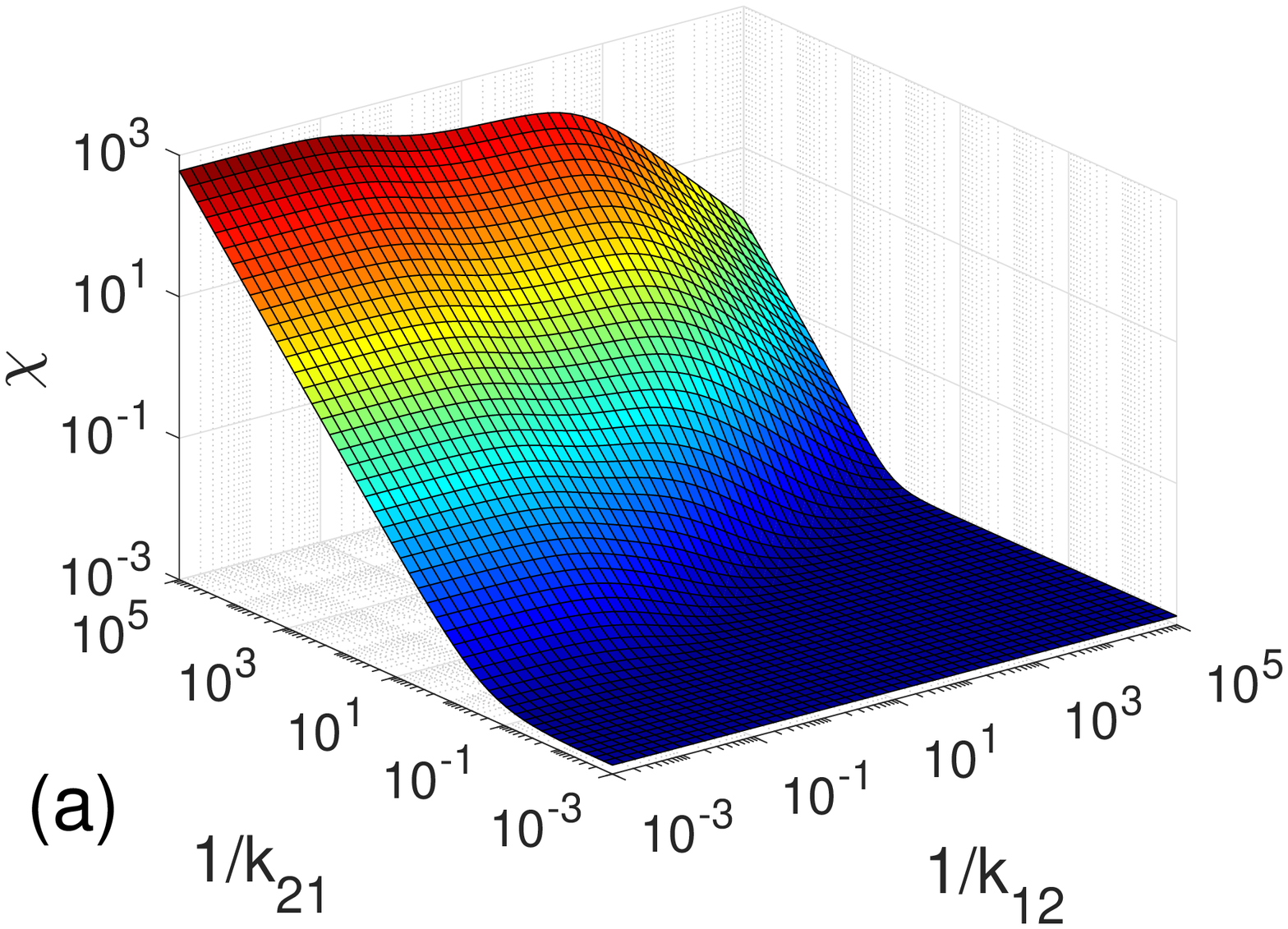} % surf_kk3a.eps}
\includegraphics[width=85mm]{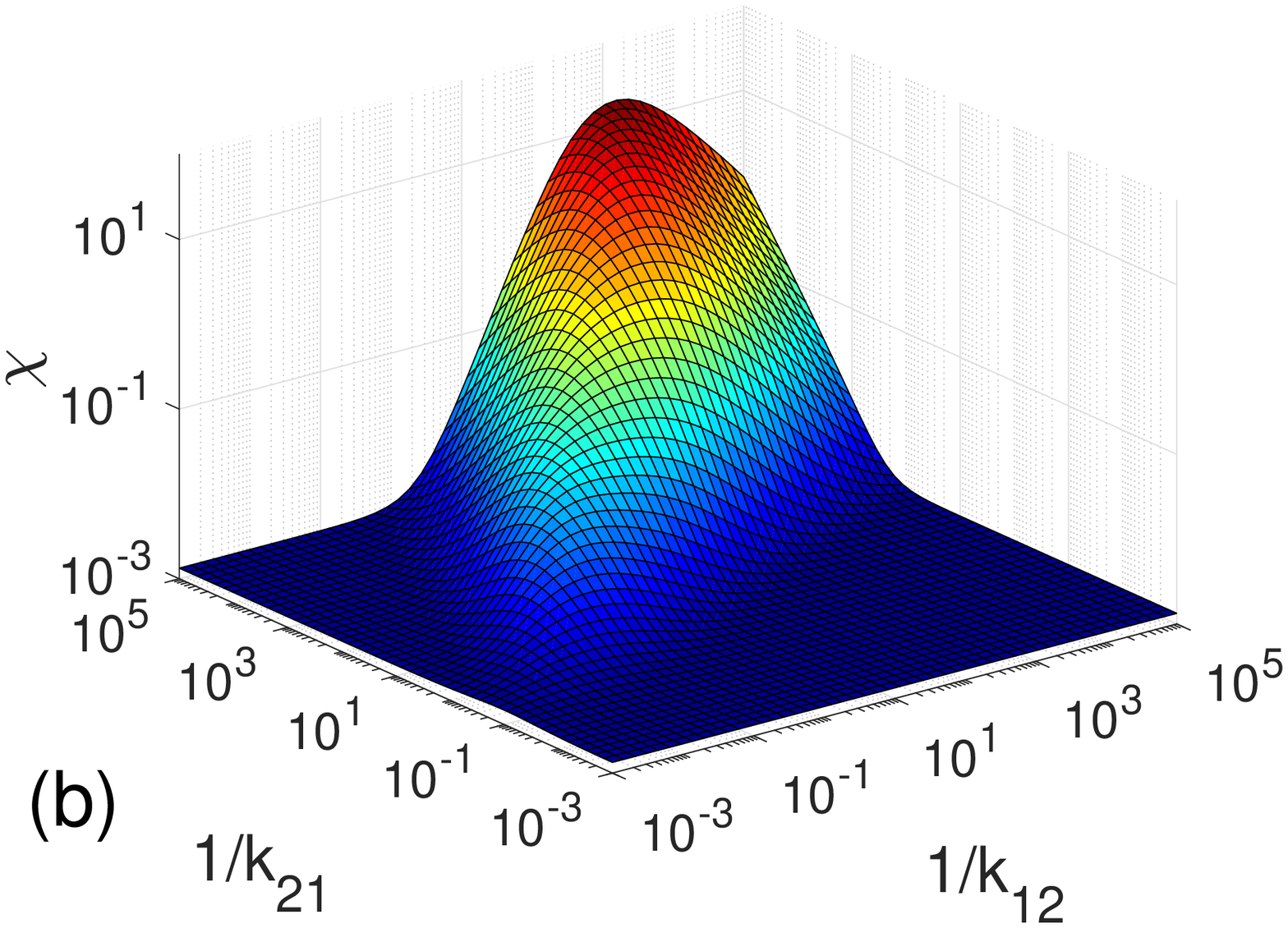} % surf_kk3b.eps}
\end{center}
\caption{
(Color online) The EB parameter $\chi$ as a function of mean residence
times $1/k_{12}$ and $1/k_{21}$, for $t = 1$, $T = 1000$, $D_1 = 1$,
and $D_2 = 0$ {\bf (a)} and $D_2 = 0.01$ {\bf (b)}.  Arbitrary units
are used.}
\label{fig:surf}
% [C1,C2] = A_Karger_chivar_surf2();
% Export 20 x 15
\end{figure}

A richer insight on the EB parameter is provided in
Fig. \ref{fig:surf} that shows the dependence on two residence times
$1/k_{12}$ and $1/k_{21}$.  In this situation, there are four relevant
time scales: $t$, $T$, $1/k_{12}$ and $1/k_{21}$ and thus various
regimes.  As previously, we fix $t = 1$ and $T = 1000$.  First, we
present in Fig. \ref{fig:surf}(a) the case of an immobile state with
$D_2 = 0$ (and $D_1 = 1$).  When $1/k_{21}$ is the smallest time
scale, the particle does not almost stay in the immobile state, and
the EB parameter is close to its value $4t/(3T)$ for normal diffusion.
When $1/k_{21}$ is getting comparable to the lag time $t$, the
switching is not fast enough any more, and the presence of the
immobile phase increases the EB parameter.  Finally, when $1/k_{21}$
increases further, the EB parameter can become much larger than $1$.
In fact, in the limit $k_{21} \to 0$ (with fixed $k_{12}$), one gets
\begin{equation}
\chi \simeq \frac{2A}{k_{12}^3(T-t)^2 t^2} \, k_{21}^{-1} + O(1),
\end{equation}
where
\begin{align*}
& A = e^{-k_{12}T}(1-e^{k_{12}t})^2 - 2(3+2k_{12}(T-t))e^{-k_{12}t} + 6\\
& + k_{12}T (3k_{12}^2t^2 + 4 - 4k_{12}t)  - 10k_{12}t + 6k_{12}^2 t^2 -4k_{12}^3t^3.
\end{align*}
One can see that the EB parameter can be made arbitrarily large by
decreasing $k_{21}$.  Indeed, a particle started in the immobile state
mainly remains in this state, whereas a particle started in the mobile
state becomes immobile with the switching rate $k_{12}$.  As a
consequence, random trajectories may have a broad distribution of
stalling periods and thus the broad distribution of TAMSD.  We note
that the other mean residence time, $1/k_{12}$, also influences the EB
parameter but it is less relevant than $1/k_{21}$.

The situation is considerably different for a particle undergoing slow
diffusion in the second state (i.e., $D_2$ is small but not strictly
zero).  Figure \ref{fig:surf}(b) shows an example for $D_2 = 0.01$.
While the behavior of the EB parameter for small $1/k_{21}$ is
expectedly similar to the former case with $D_2 = 0$, there is
significant difference for large $1/k_{21}$.  First, the values of the
EB parameter, which can still be large, are much smaller than those
shown in Fig. \ref{fig:surf}(a).  Second, the EB parameter exhibits a
maximum as a function of the mean residence time $1/k_{12}$ for a
fixed $1/k_{21}$.  When $1/k_{12}$ is small (with $1/k_{21}$ large),
the particle stays most of the time in the second state and thus
undergoes almost normal diffusion with diffusivity $D_2$, so that the
EB parameter again reaches the small value $4t/(3T)$.  This is a
significant difference with respect to the case $D_2 = 0$.

\section{Discussion}
\label{sec:discussion}

When the measurement time $T$ is comparable to or smaller than the
residence times $1/k_{12}$ and $1/k_{21}$, the particle does not have
enough time to probe the phase space.  An individual random trajectory
of the particle is thus not representative of the ensemble, suggesting
an apparent weak ergodicity breaking.  In particular, the EB parameter
can be of order of $1$ and even much larger, particularly if the
diffusion coefficient in one of the states is very small (or zero).
As a consequence, one can expect a large spread of TAMSD curves
evaluated from individual trajectories.  This feature, which is often
observed in experiments (see, e.g.,
\cite{Golding06,Jeon11,Mosqueira18}), was often interpreted as an
indication of non-ergodicity and thus attributed to CTRW as a basic
non-ergodic model.  Here we showed that the ergodic two-state model
can lead to similar features.

\subsection{Spread of TAMSD curves}

In order to illustrate the spread of TAMSD curves, we perform Monte
Carlo simulations.  For a prescribed set of parameters $D_1$, $D_2$,
$k_{12}$ and $k_{21}$, we set the trajectory duration $T = 1000$
(i.e., the number of steps) and generate $T$ random centered Gaussian
increments with unit variance.  We also generate a sequence of
successive residence times in two alternating states according to the
exponential laws with the rates $k_{12}$ and $k_{21}$, that results in
a random sequence of state variables $i_1, i_2, \ldots$ (each $i_n$
taking values $1$ or $2$).  All the increments are rescaled by
$\sqrt{2D_{i_n}}$ and then cumulatively summed up to produce a random
trajectory, from which the TAMSD is computed for all lag times from
$1$ to $T-1$ by discretizing Eq. (\ref{eq:TAMSD}).  This computation
is repeated $10\,000$ times to obtain a reliable statistics.  Fixing
$D_1 = 1$, we are left with $D_2$, $k_{12}$ and $k_{21}$ as the major
parameters.

\begin{figure}
\begin{center}
\includegraphics[width=85mm]{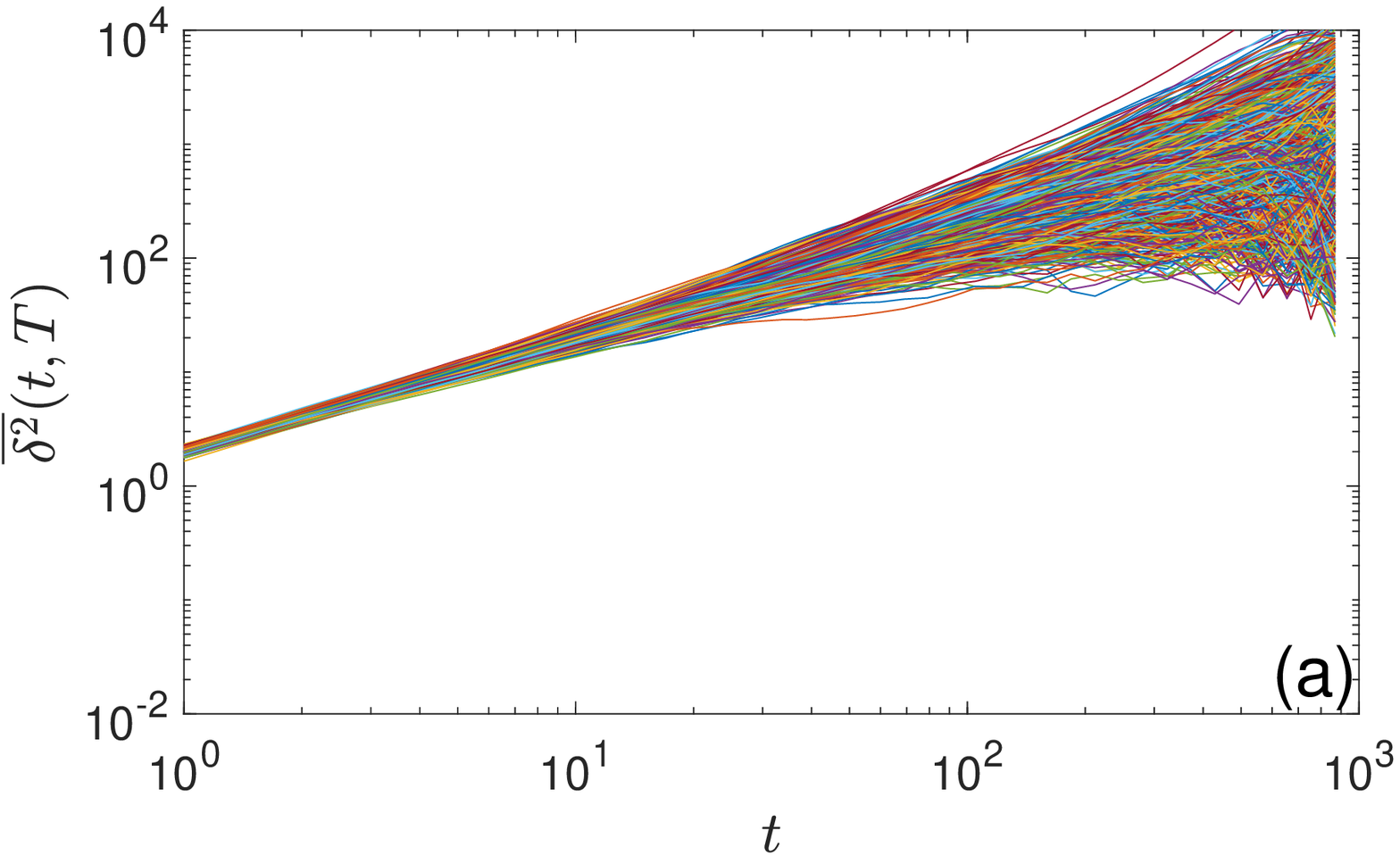} % TAMSD_normal_.eps}
\includegraphics[width=85mm]{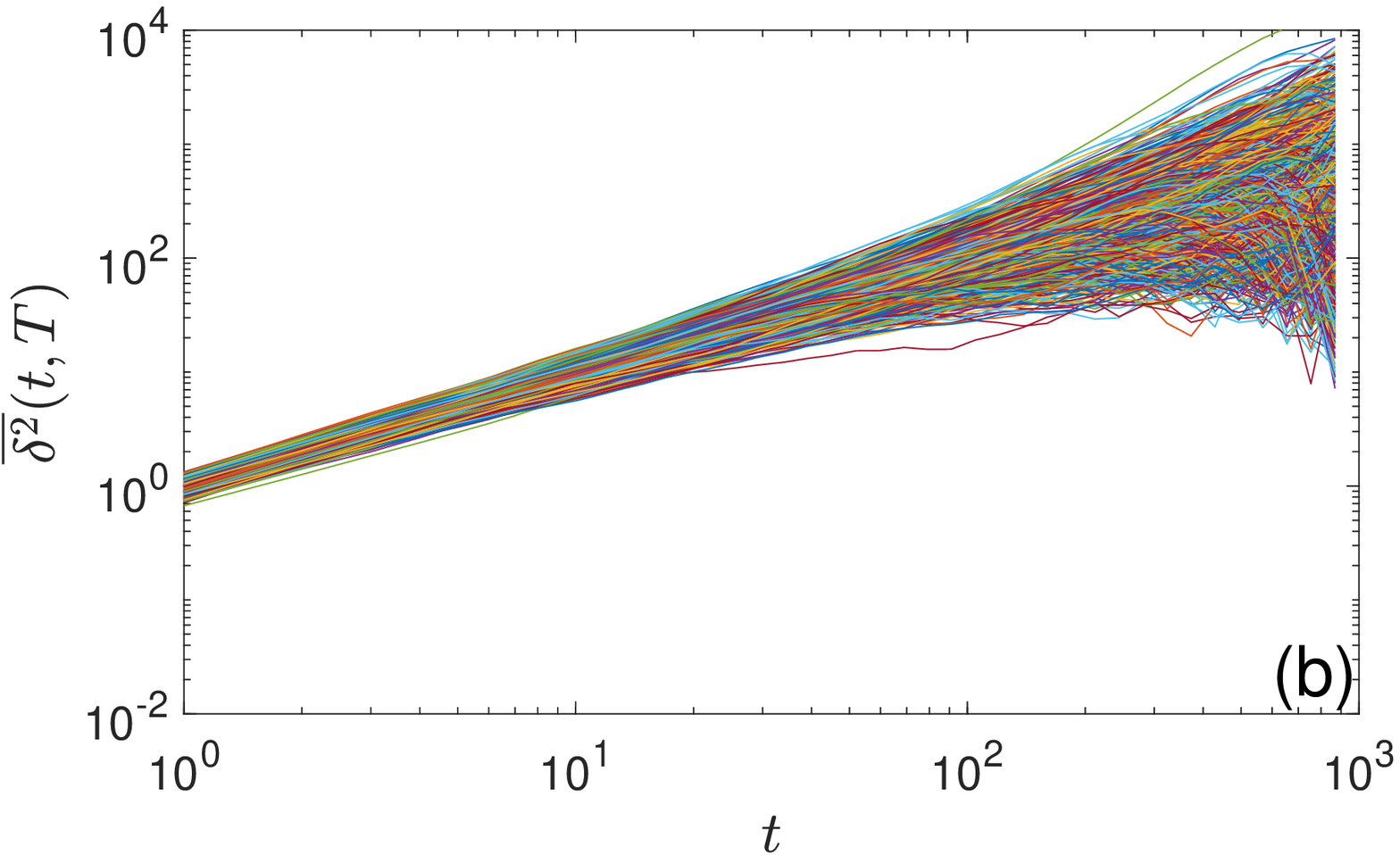} % TAMSD_narrow_.eps}
\includegraphics[width=85mm]{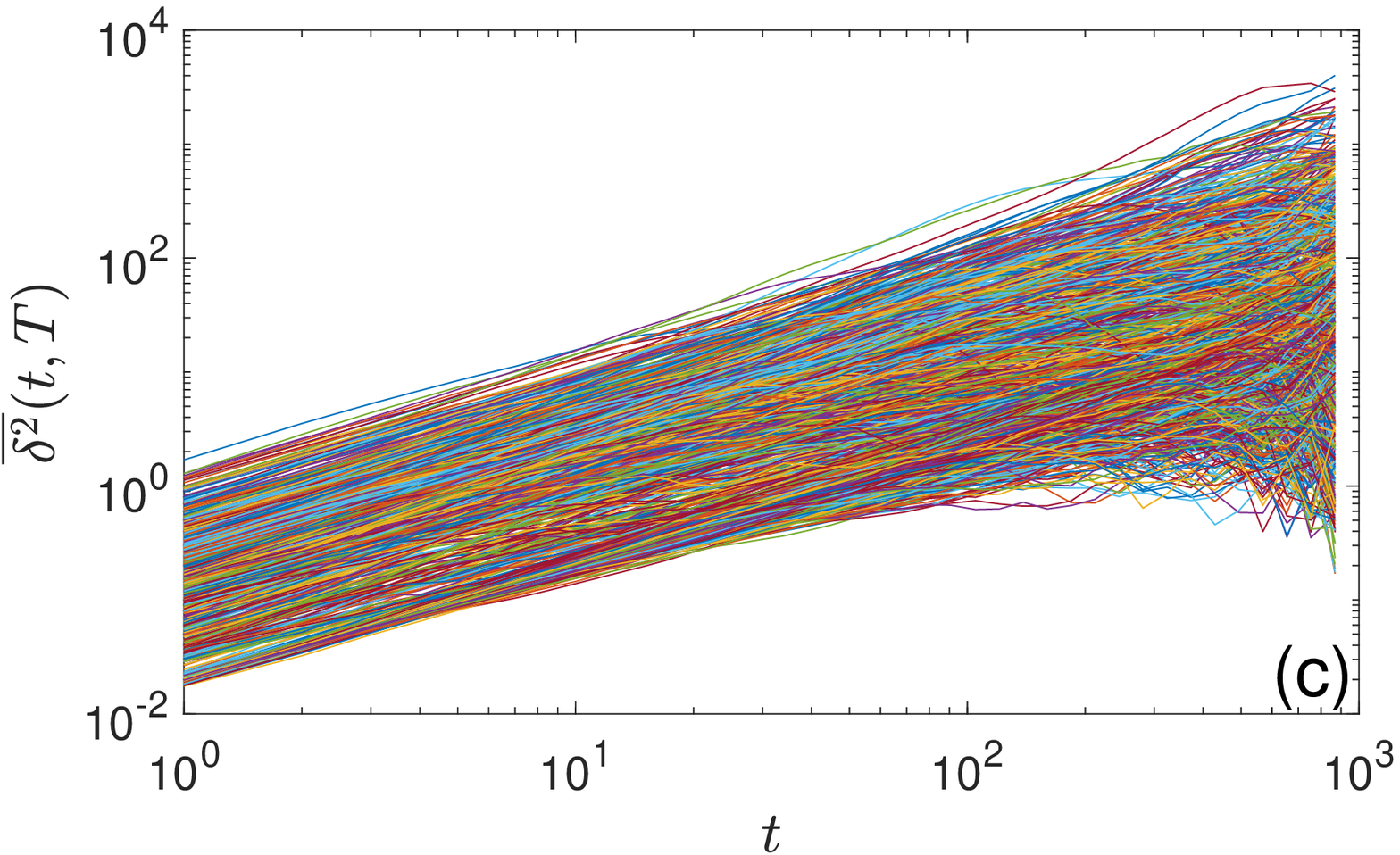} % TAMSD_broad_.eps}
\end{center}
\caption{
(Color online) Spread of TAMSD curves obtained from a thousand of
simulated trajectories of length $T = 1000$, with $D_1 = 1$, $p_i =
p_i^{\rm eq}$, and three sets of parameters: {\bf (a)} $D_2 = 1$
(normal diffusion); {\bf (b)} $D_2 = 0.01$ and $k_{12} = k_{21} =
10^{-1}$; and {\bf (c)} $D_2 = 0.01$, $k_{12} = 10^{-2}$ and $k_{21} =
10^{-3}$.  Arbitrary units are used.}
\label{fig:spread}
% [dX,X,MSD,tt] = A_Karger_simu_fig;
\end{figure}

Figure \ref{fig:spread} illustrates the spread of TAMSD curves for
three sets of parameters.  The first set with $D_2 = 1$ corresponds to
normal diffusion (here, switching rates do not matter as $D_1 = D_2$).
In this case, TAMSD curves are close to each other at small lag times
and then getting more spread at larger lag times because the time
average becomes less and less efficient (Fig. \ref{fig:spread}a).  In
the second set (with $D_2 = 0.01$, $k_{12} = k_{21} = 0.1$), the mean
residence times $1/k_{12}$ and $1/k_{21}$ are chosen to be much
smaller than the measurement time $T$.  As switching is rapid enough,
TAMSD curves remain close to each (Fig. \ref{fig:spread}b), as for
normal diffusion.  In the third set of parameters, we keep $D_2 =
0.01$ but decrease both switching rates: $k_{12} = 10^{-2}$ and
$k_{21} = 10^{-3}$.  While $1/k_{12}$ is still much smaller than $T$,
$1/k_{21}$ is equal to $T$ that leads to a wide spread of TAMSD curves
(Fig. \ref{fig:spread}c), in agreement with large values of the EB
parameter in this case.  Similar spreads were observed in
single-particle tracking experiments in living cells (see, e.g.,
\cite{Golding06,Jeon11,Mosqueira18}).

\begin{figure}
\begin{center}
\includegraphics[width=85mm]{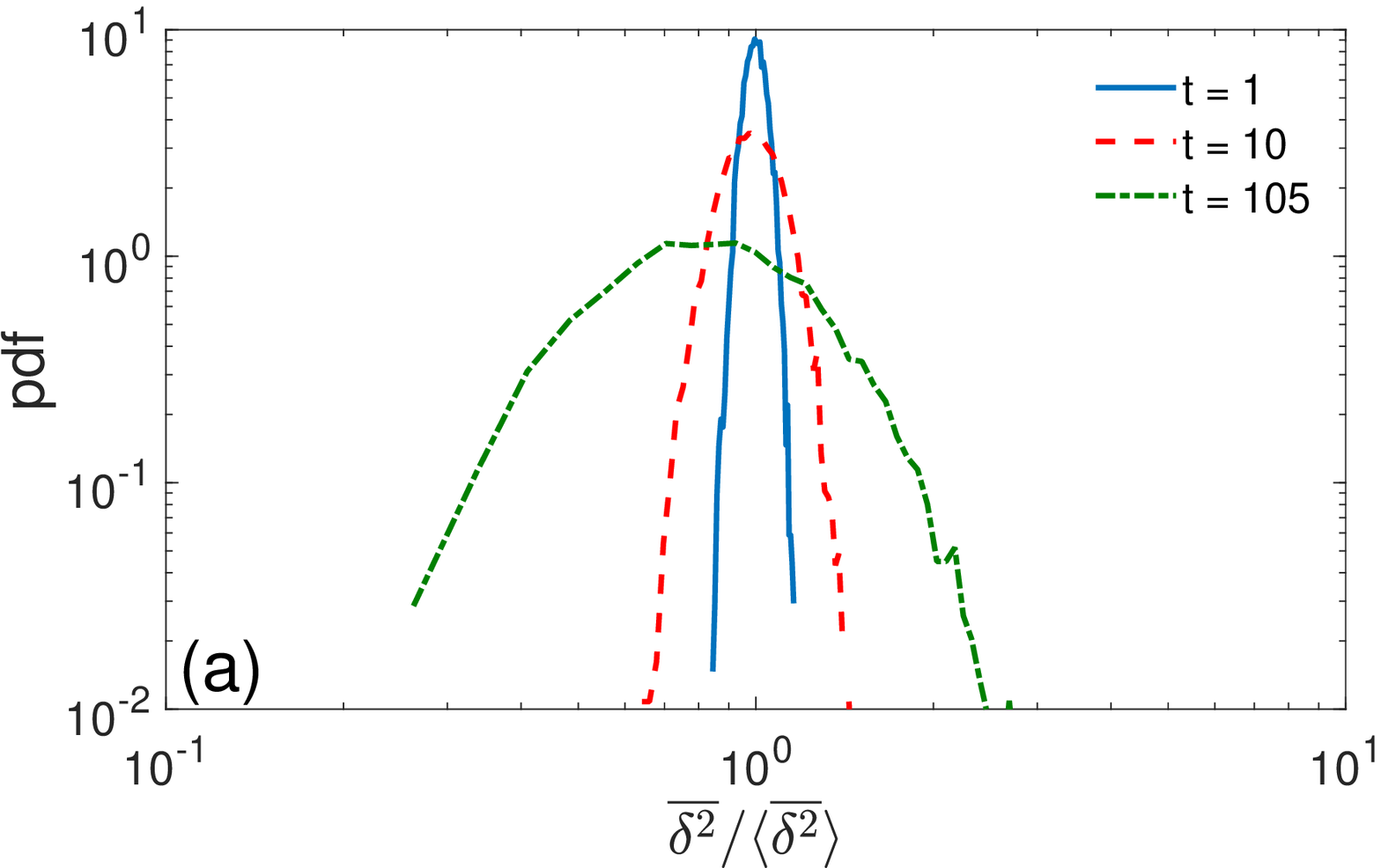} % TAMSD_dist_normal_.eps}
\includegraphics[width=85mm]{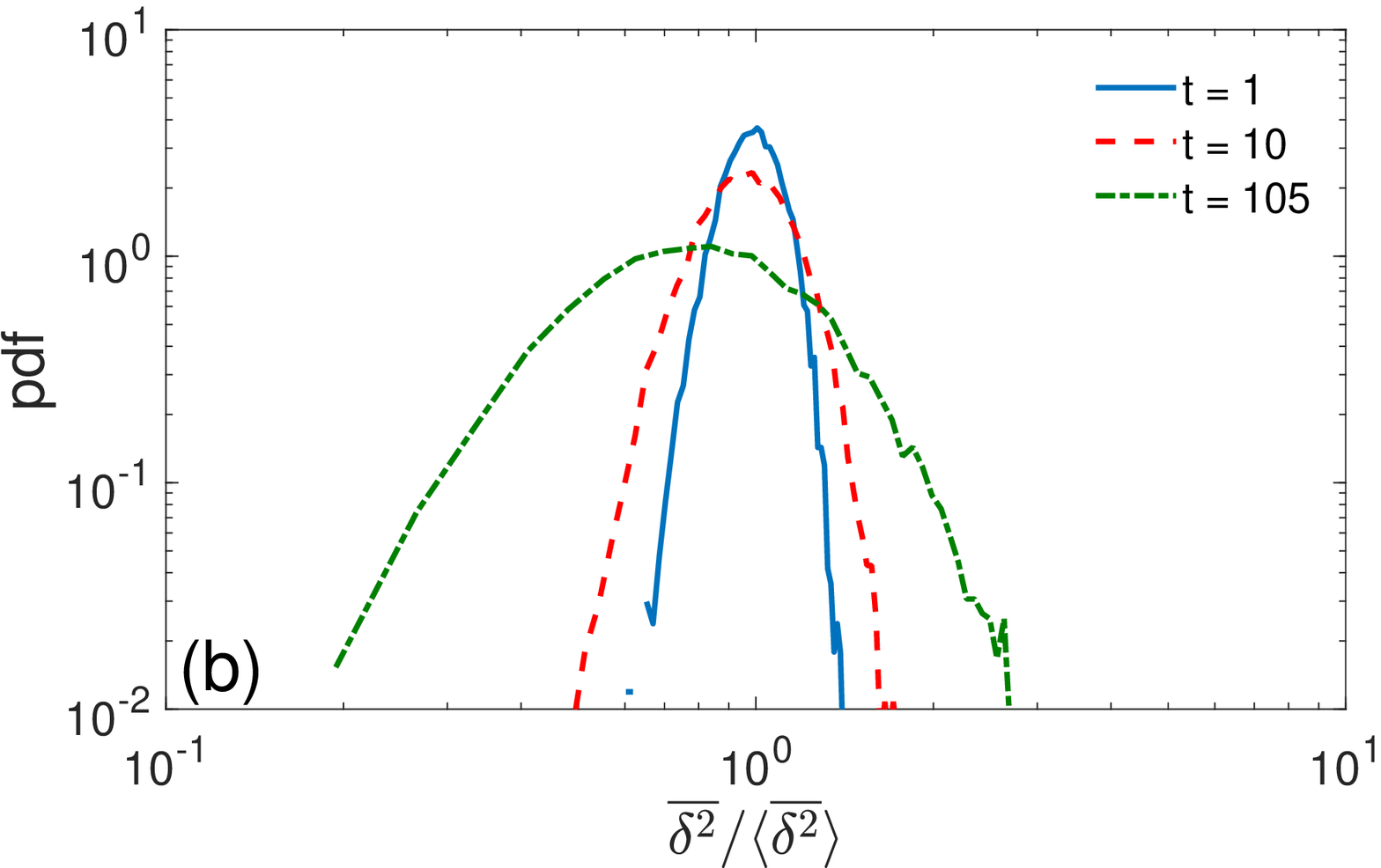} % TAMSD_dist_narrow_.eps}
\includegraphics[width=85mm]{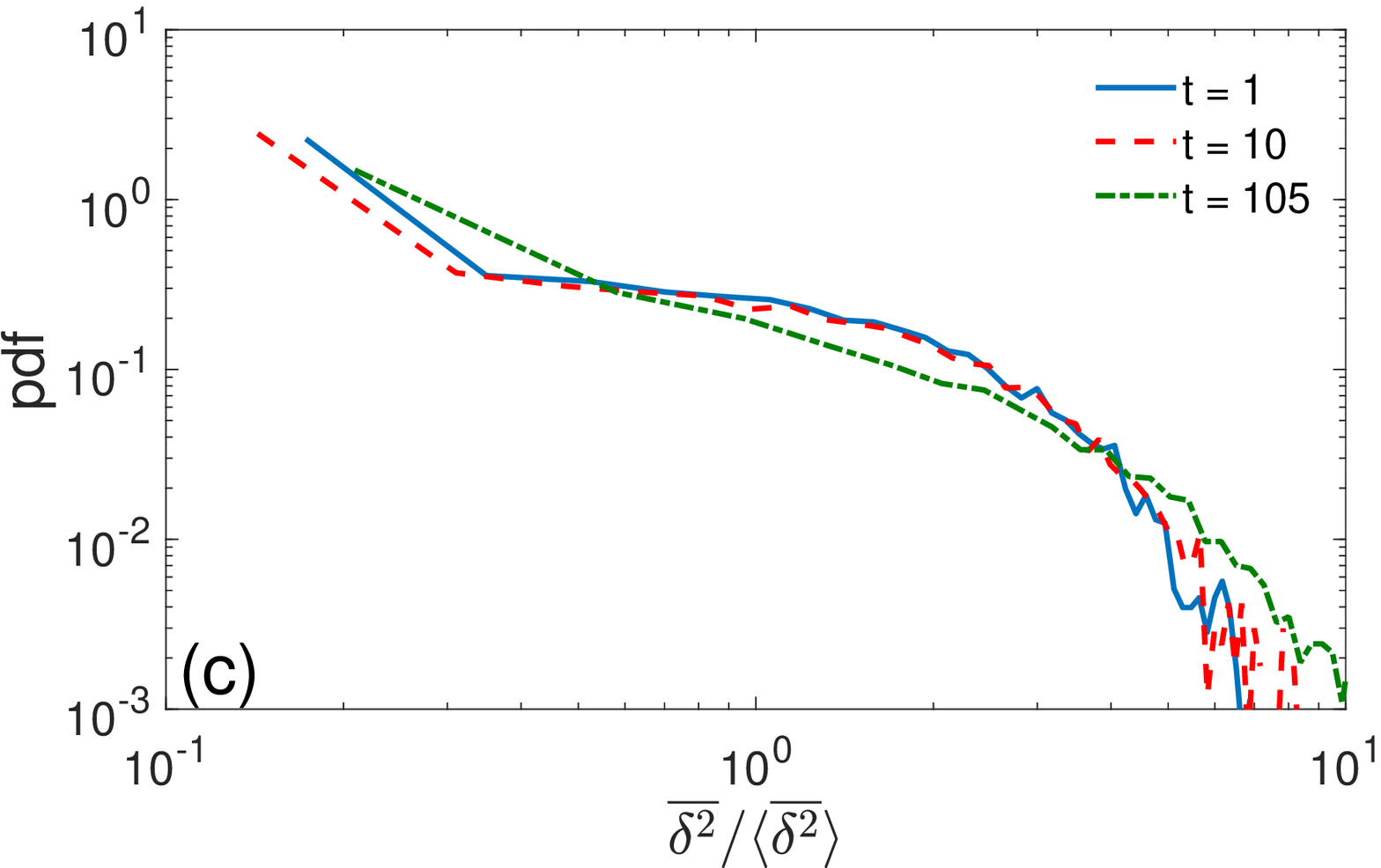} % TAMSD_dist_broad_.eps}
\end{center}
\caption{
(Color online) Empirical distribution of TAMSD,
$\overline{\delta^2}(t,T)$, normalized by its mean value $\langle
\overline{\delta^2}(t,T) \rangle$, obtained from 10\,000 simulated
trajectories of length $T = 1000$, with $D_1 = 1$, $p_i = p_i^{\rm
eq}$, and three sets of parameters: {\bf (a)} $D_2 = 1$ (normal
diffusion); {\bf (b)} $D_2 = 0.01$ and $k_{12} = k_{21} = 10^{-1}$;
and {\bf (c)} $D_2 = 0.01$, $k_{12} = 10^{-2}$ and $k_{21} = 10^{-3}$.
Arbitrary units are used.}
\label{fig:distrib}
% [dX,X,MSD,tt] = A_Karger_simu_fig2;
\end{figure}

Another way of presenting the spread consists in plotting the
distribution of TAMSD for a fixed lag time.  The distribution of TAMSD
for ergodic Brownian motion and other Gaussian processes was studied
in \cite{Grebenkov11,Andreanov12,Sikora17,Gajda18}, while the analysis
of its asymptotic behavior for non-ergodic CTRW was initiated in
\cite{He08,Lubelski08} (see also reviews \cite{Burov11,Metzler14}).  As
our computation for two-state switching diffusion is limited to the
first two moments, we show in Fig. \ref{fig:distrib} the empirical
distribution of TAMSD obtained from simulated trajectories for the
same three sets of parameters.  The empirical distributions are
presented for three lag times: $t = 1$, $t = 10$ and $t = 105$.  As
expected, the distribution is getting larger with the lag time,
reflecting less and less efficient time averaging.  For the second set
of parameters, the distributions are close to that for normal
diffusion (compare Figs. \ref{fig:distrib}a and \ref{fig:distrib}b).
In contrast, the distribution for the third set of parameters is much
broader and almost does not depend on the lag time.  This is the
characteristic feature of non-ergodic dynamics (e.g., the distribution
of TAMSD for CTRW is also broad).

To complete this discussion, we outline a subtle difference in
the behavior of the EB parameter between a continuous-time process and
its discrete-time approximation.  Let us first discuss Brownian
motion, which is often approximated by random walks on a lattice or by
a sequence of Gaussian jumps.  While such approximations are known to
converge to Brownian motion (see, e.g., \cite{Morters,Grebenkov16}),
some functionals involving integrals along sample trajectories of
these processes can be different.  So, the variance of the TAMSD for
continuous-time Brownian motion, given by the first term in
Eqs. (\ref{eq:varchi1}, \ref{eq:varchi2}), is different from the
variance of the TAMSD for a discrete-time random walk with Gaussian
jumps, first derived by Qian {\it et al.} \cite{Qian91} and later
analyzed in \cite{Grebenkov11,Grebenkov13}.  In particular, when $t
\ll T$, the leading term of the EB parameter is $4t/(3T)$ in the
continuous case, and $2t/T$ in the discrete case.  While the scaling
form, $t/T$, is identical for both cases, the numerical prefactor is
different.  This difference is related to the fact that an erratic
trajectory of Brownian motion between two successive discrete times is
replaced by a jump in the discrete case.

There is a similar distinction between the continuous-time switching
diffusion, studied in Sec. \ref{sec:theory}, and discrete-time Monte
Carlo simulations presented here.  In particular, the variance of the
TAMSD, computed via the exact formulas (\ref{eq:varchi1},
\ref{eq:varchi2}), would differ from its numerical estimation by Monte
Carlo simulations.  In spite of this subtle difference between
continuous-time and discrete-time processes, the qualitative
conclusion on distinct diffusivity states and insufficient measurement
time as eventual causes of the apparent weak ergodicity breaking
remains valid, as confirmed by numerical results of this section.  The
exact computation of the EB parameter for a discrete-time switching
diffusion by adapting the general technique from \cite{Grebenkov11}
presents an interesting perspective for future research.

\subsection{Extension to multi-state models}
\label{sec:extension1}

The above computation can be formally extended to a switching model
with $J$ states, which is characterized by a set of diffusion
coefficients $D_i$ and switching rates $k_{ij}$.  The propagators
satisfy for each $i_0 = 1,\ldots,J$:
\begin{equation}
\frac{\partial}{\partial t} P_{i,i_0} = D_i \frac{\partial^2}{\partial x^2} P_{i,i_0} + \sum\limits_{j=1}^J k_{ji} P_{j,i_0} ,
\end{equation}
where $k_{ii} \equiv - \sum\limits_{j\ne i} k_{ij}$.  The formula
(\ref{eq:Pmultiple}) for the marginal propagator for multiple
successive points in the Fourier space remains practically unchanged:
\begin{eqnarray}  \label{eq:Pmultiple_K}
&& {\mathcal P}_n(q_n,t_n-t_{n-1};\,   \ldots;   q_2,t_2-t_1;\,  q_1,t_1) \\  \nonumber
&& = \left(\begin{array}{c} 1 \\ 1 \\ ... \\ 1 \\ \end{array} \right)^\dagger
\Pq(q_n,t_n-t_{n-1})   \ldots  \Pq(q_1,t_1) \left(\begin{array}{c} p_1 \\ p_2 \\ ... \\ p_J \\ \end{array} \right),
\end{eqnarray}
where $p_i$ is the probability of starting in the state $i$, $\Pq(q,t)
= \exp(-\M_q t)$, and $\M_q$ is now the $J \times J$ matrix of the
form $\M_q = q^2 \D - \K^\dagger$, with $(\D)_{ij} =
\delta_{ij} D_i$ and $(\K)_{ij} = k_{ij}$.  As the eigenvalues
$\gamma_q$ of the matrix $\M_q$ are determined as the zeros of the
polynomial
\begin{equation}  \label{eq:det}
\det(\gamma {\bf I} - \M_q) = 0 ,
\end{equation}
there is no explicit formula in general.  However, as we are
interested in evaluating the limit $q\to 0$ to get the mean square
displacement and other related quantities, one can apply the standard
perturbation theory by treating $q^2 \D$ as a perturbation of the
matrix $\M_0 = -\K^\dagger$.  As the matrix $\M_0$ is neither
symmetric, nor invertible, the analysis is more involved.  While
analytical calculations for a general multi-state model seem
challenging, numerical computations of the propagator and related
quantities (such as the mean square displacement) are efficient due to
the matrix form (\ref{eq:Pmultiple_K}) when the number of states is
not too high (say, below a thousand).
It is also worth noting that multi-state switching diffusions can be
seen as a discrete version of diffusing diffusivity models
\cite{Chubynsky14,Jain16,Chechkin17,Lanoiselee18a,Lanoiselee18b,Sposini18,Sposini19},
in which the diffusivity $D_t$ of a tracer changes continuously in
time (see \cite{Grebenkov19} for details).

\subsection{Extensions to switching diffusion in bounded domains}
\label{sec:extension2}

While we focused on one-dimensional diffusion, the above computation
can be carried on for more sophisticated processes.  Here, we briefly
mention switching diffusion in an arbitrary bounded domain $\Omega
\subset \R^d$, for which the propagator for multiple successive
points, $P(\x_n,i_n,t_n;\, \x_{n-1},i_{n-1},t_{n-1}; \,
\ldots; \x_0,i_0,0)$, and its average over the state variables $i_k$,
$P(\x_n,t_n;\, \x_{n-1},t_{n-1}; \, \ldots; \x_0,0)$, can be expressed
in analogy to Eqs. (\ref{eq:multiple_prop1}, \ref{eq:multiple_prop2}).
For instance, Eq. (\ref{eq:multiple_prop2}) becomes
\begin{eqnarray}  \nonumber
&& P(\x_n,t_n;\, \x_{n-1},t_{n-1}; \, \ldots; \x_0,0)  
 = \sum\limits_{k_1,\ldots,k_n} u_{k_n}(\x_n)  \\  \nonumber
&& u_{k_n}^*(\x_{n-1}) u_{k_{n-1}}(\x_{n-1}) u_{k_{n-1}}^*(\x_{n-2}) \ldots u_{k_1}(\x_1) u_{k_1}^*(\x_0) \\
&& \times {\mathcal P}_n(\sqrt{\lambda_{k_n}},t_n-t_{n-1}; \, \ldots; \sqrt{\lambda_{k_1}},t_1) ,
\end{eqnarray}
where ${\mathcal P}_n$ is given by Eq. (\ref{eq:Pmultiple}) (or
Eq. (\ref{eq:Pmultiple_K}) in a multi-state case), while $\lambda_n$
and $u_k(\x)$ are the eigenvalues and $L_2$-normalized eigenfunctions
of the Laplace operator $\Delta$, satisfying $\Delta u_k(\x) +
\lambda_k u_k(\x) = 0$ in $\Omega$.  Boundary conditions determine the
properties of the boundary in a standard way \cite{Redner}: Neumann
condition incorporates passive impermeable walls whereas Dirichlet or
Robin conditions allow one to describe diffusion-controlled reactions
and the related first-passage time statistics, see
\cite{Lanoiselee18b} (note that ${\mathcal P}_1(q,t)$ corresponds to
$\Upsilon(t;q^2)$ in \cite{Lanoiselee18b}).  For instance, one can
investigate diffusion-controlled reactions in presence of buffers that
can reversely bind the diffusing molecule and thus affect its
mobility.  We emphasize that the above derivation requires that
switching modifies only the diffusivity of the particle but does not
change other properties (e.g., reactivity).  In this case, restricted
diffusion in each state is characterized by the same set of Laplacian
eigenmodes.  In contrast, this extension is not applicable to other
intermittent processes \cite{Benichou11b} such as, e.g.,
surface-mediated diffusion
\cite{Benichou10,Benichou11,Rojo11,Rupprecht12a,Rupprecht12b},
two-channel diffusion with switching reactivity \cite{Godec17} or
switching boundary conditions \cite{Bressloff15,Lawley16}.

From the propagator, one can compute the mean square displacement, as
well as the mean and the variance of the TAMSD.  However, these
formulas include multiple infinite sums over eigenvalues and thus
remain formal, so we do not present these results.

\section{Conclusion}
\label{sec:conclusion}

We revisited the two-state switching diffusion model from a
single-particle tracking perspective.  The exact formulas for the mean
and the variance of the TAMSD were derived, in order to investigate
the behavior of the ergodicity breaking parameter.  When the mean
residence times $1/k_{12}$ and $1/k_{21}$ are small as compared to the
measurement time $T$, switching is fast enough for a particle to probe
both states so that a single trajectory is representative of the
ensemble.  In particular, the distribution of the TAMSD is narrow to
allow for an accurate estimation of the mean diffusivity $\bar{D}$.
Roughly speaking, the two-state switching diffusion looks as normal
diffusion with diffusivity $\bar{D}$.  In contrast, when $1/k_{12}$
and/or $1/k_{21}$ are comparable to or larger than $T$, an individual
trajectory is not representative of the ensemble, the EB parameter may
exceed $1$, and the distribution of the TAMSD is broad.  As a
consequence, TAMSD curves exhibit a significant spread, in agreement
with experimental observations in living cells.  While the two-state
switching diffusion is ergodic (when $k_{12}$ and $k_{21}$ are
strictly positive), the analysis of individual trajectories may
indicate non-ergodic features.  As this fictitious non-ergodicity is
the mere consequence of insufficient measurement time $T$, we called
it ``apparent weak ergodicity breaking''.  A similar behavior was
observed for CTRW with an exponential cut-off: when the measurement
time is much smaller than the cut-off time, the trajectory ``looks''
as that of usual non-ergodic CTRW but when the measurement time grows
and exceeds the cut-off time, the non-ergodic features disappear
\cite{Lanoiselee16}.  The relative simplicity of switching diffusions
presents their advantage as potential models for describing some
intracellular and on-membrane transport \cite{Sungkaworn17,Weron17}.
Moreover, these are natural models for describing the effect of
buffers that can reversely bind the diffusing particle and thus affect
its mobility.  More generally, our results illustrate that a broad
spread of TAMSD curves, often observed in experiments, is not
necessarily a signature of weak ergodicity breaking.  It may just be a
consequence of the insufficient duration of acquired trajectories.
Statistical tests of ergodicity breaking need to be systematically
performed in such situations \cite{Magdziarz11,Lanoiselee16}.

\begin{acknowledgments}

DG acknowledges the financial support by French National Research
Agency (ANR Project ANR-13-JSV5-0006-01).  

\end{acknowledgments}

\end{document}